\documentstyle[12pt]{article}
\begin{document}
\title{A Brief Note on Discrete Space Effects}
\author{B.G. Sidharth\\Centre for Applicable Mathematics \& Computer Sciences\\
B.M. Birla Science Centre, Hyderabad 500 063}
\date{}
\maketitle
\begin{abstract}
We discuss briefly discrete space effects and show that this leads to space
reflection asymmetry and also a minor modification of Einstein's energy-mass
formula.
\end{abstract}
Snyder\cite{r1} had shown how it is possible to consider discrete space time
consistently with the Lorentz transformation. Discrete space time continues to
receive attention over the years from several scholars (Cf. for example \cite{r2,r3,r4,r5,r6,r7}
for details). Indeed in the Dirac's relativistic theory of the electron, this
discretization is evident - averages over the Compton scale are required to
eliminate Zitterbewegung effects and non Hermitian position operators and to
recover meaningful physics\cite{r8,r9} - and that includes special relativity.
It is in this light that a recent formulation of
an electron in terms of a Kerr-Newman metric becomes meaningful\cite{r9,r10,r11}
and further this leads to a meaningful, if phenomenological mass spectrum
\cite{r12,r13,r14}.\\
All this pleasingly dovetails with the fact that if the minimum space time
cut offs are taken at the Compton scale, then we have a non commutative
geometry viz.,
\begin{equation}
[x,y] = 0 (l^2)\label{e1}
\end{equation}
and similar equations, and which further, leads directly to the Dirac equation
(Cf.\cite{r15} for details). This can be easily seen from the fact that, given
(\ref{e1}), the usual infinitessimal coordinate shift in Minkowski space,
is,
\begin{equation}
\psi' (x_j) = [1 + \imath \epsilon (\imath \epsilon_{ljk} x_k \frac{\partial}
{\partial x_j}) + 0 (\epsilon^2)] \psi (x_j)\label{e2}
\end{equation}
The choice
$$
t =  \left(\begin{array}{l}
        1 \quad 0 \\ 0 \quad -1 \\
        \end{array}\right), \vec x =
   \ \  \left(\begin{array}{l}
         0 \quad \vec \sigma \\ \vec \sigma \quad 0 \\
       \end{array}\right)$$
provides a representation for the coordinates, as can be easily verified and
then from (\ref{e2}) we recover the Dirac equation.\\
If on the other hand terms $\sim l^2$ are neglected, then (\ref{e1}) gives the
usual commutation relations of Quantum Theory. Discrete
space time is therefore a higher order correction to usual Quantum Theory.\\
In this context it has been pointed out that the discrete time provides an
explanation for the puzzling Kaon decay which violates time reversal
symmetry and also the decay of a pion into an electron and a positron\cite{r16,r8,r9,r12,r13}.\\
We now observe that from an intuitive point of view space or time reversal
symmetries based on space time points theory cannot be taken for granted if
space time is discrete. This can immediately be seen from (\ref{e1}): If we
retain terms $\sim l^2$, then there is no invariance under space reflections.\\
Indeed in the same vein, as discussed earlier\cite{r9,r10}, the fact that the
Compton wavelength of the nearly massless neutrino is very large provides an
explanation of its handedness.\\
We finally point out another $0(l^2)$ effect,
which can be demonstrated in a simple way by invoking the derivation of the
wave equation of a particle by replacing the continuum by a set
of "lattice points"\cite{r9,r17}. In
this case we have an equation like
$$Ea(x_n) = E_0a(x_n) - Aa(x_n + b) - Aa(x_n - b).$$
where $b$ is the distance between successive "lattice points". This leads to
\begin{equation}
E = E_0 - 2A cos kb.\label{e3}
\end{equation}
We can choose the zero of energy in such a way that when $b \to 0$, we have $E = 2A$
which is then identified with the rest energy $mc^2$ of the particle(Cf.\cite{r9} for
details).\\
However if we do not neglect terms $\sim b^2 = l^2$, then we have from (\ref{e3})
$$\left| \frac{E}{mc^2} - 1\right| \sim 0 (l^2)$$
It must be reiterated that when terms $\sim l^2$ are neglected, we recover
the usual theory. Finally, the above conclusions are true with minor modifications
in case the minimum cut off is non-zero but not the Compton wavelength.


\begin{thebibliography}{99}
\bibitem {r1} Snyder, H.S., Phys. Rev. {\bf 71 } (1), 1947, 38ff.
\bibitem {r2} Kadyshevskii, V.G., Soviet Physics Doklady {\bf 7 } (11), 1963, 1030.\\
\bibitem {r3} Kadyshevskii, V.G., Soviet Physics Doklady {\bf 7 } (12), 1963, 1138ff.
\bibitem {r4} Wolf C., Nuovo. Cim. B 109 (3), 1994, 213.
\bibitem {r5} Bombelli, L., Lee, J., Meyer, D., Sorkin, R.D., Phys. Rev. Lett.
59, 1987, 521.
\bibitem {r6} Caldirola, P., Lettere Al Nuovo Cimento, Vol.16, N.5, 1976, 151ff.
\bibitem {r7} Lee, T.D., Phys. Lett. 12 (2B), 1983, 217.
\bibitem {r8} Dirac, P.A.M., "Principles of Quantum Mechanics", Clarendon
Press, Oxford, 1958.
\bibitem {r9} Sidharth, B.G., Ind. J.  Pure and Applied Phys., 35 (7), 1997, 456.
\bibitem {r10} Sidharth, B.G., IJMPA, 13 (15), 1998, 2599. Also xxx.lanl.gov quant-ph
9808031.
\bibitem {r11} Sidharth, B.G., Gravitation and Cosmology {\bf 4 } (2) (14), 158ff (1998) and
references therein.
\bibitem {r12} Sidharth, B.G., Mod.Phys. Lett. A., Vol. 12 No.32, 1997, pp2469-2471.
\bibitem {r13} Sidharth, B.G., Mod.Phys. Lett. A., Vol. 14 No. 5, 1999, pp387-389.
\bibitem {r14} Sidharth, B.G., and Lobanov, Yu Yu, Proceedings of Frontiers of Fundamental Physics,
(Eds.)B.G. Sidharth and A. Burinskii, Universities Press, Hyderabad, 1999.
\bibitem {r15} Sidharth, B.G., Chaos, Solitons and Fractals, 11 (8), 2000,
1269-1278.
\bibitem {r16} Sidharth, B.G., Chaos, Solitons and Fractals, 11 (8), 2000, 1171-1174.
\bibitem {r17} Feynman, R.P., "The Feynman Lectures on Physics", (Vol.2),
Addison-Wesley, Mass, 1965.
\end{thebibliography}
\end{document}